\documentclass[11pt,twoside]{article}
\input ./journals.def
\usepackage{asp2004}
\usepackage{psfig}
\newcommand{\YDC}{\mbox{YoDeC}}
\newcommand{\YDCs}{\mbox{YoDeCs}}
\newcommand{\msun}{\mbox{${\rm M}_\odot$}}
\newcommand{\rsun}{\mbox{${\rm R}_\odot$}}
\newcommand{\rcore}{\mbox{${\rm r}_{\rm c}$}}

\def\apgt{\ {\raise-.5ex\hbox{$\buildrel>\over\sim$}}\ }
\def\aplt{\ {\raise-.5ex\hbox{$\buildrel<\over\sim$}}\ }

\begin{document}
\title{Simulating young star clusters with primordial binaries}

\author{Simon F. Portegies Zwart}
\affil{Astronomical Institute `Anton Pannekoek', 
       University of Amsterdam, Kruislaan 403, the Netherlands and
       Institute for Computer Science, University of Amsterdam, Kruislaan 403 
}
\author{Stephen L.\ W.\ McMillan}
\affil{Department of Physics, Drexel University, Philadelphia, PA 19104, USA}

\begin{abstract}
We simulate a cluster of 144179 stars including 13107 primordial hard
binaries (10\% of the total number of single stars and binary centers
of mass), using direct integration of the equations of motion of all
stars and binaries and incorporating the effects of stellar and binary
evolution. The initial conditions are representative of young dense
star clusters in the Local Group and other nearby galaxies like the
Antennae and M82.
We find that the early phase of core collapse, driven by mass
segregation, is not appreciably delayed by the presence of a large
number of hard binaries.  By the end of the simulation, at an age of
115\,Myr, the cluster radius has expanded by about a factor of two.
This may be explained as adiabatic expansion driven by the loss (via
stellar evolution) of $\sim 40$\% of the initial total mass.  Binary
dynamics apparently has little effect on the early cluster expansion.
During the evolution, the total binary fraction drops at a roughly
constant rate of $\sim0.01$\% per Myr. The fraction of very hard
binaries, however increases at about 0.025\% per Myr. By the end of
the simulation the cluster contains 37 binaries containing at least
one black hole; roughly half (17) of these contain two black holes.

\end{abstract}

\section{Introduction}
Young and dense star clusters (\YDCs) are a relatively new class of
object, containing about $10^4$--$10^6$ stars and having ages less
than about 10\,Myr and densities exceeding a few times $10^5$
stars/pc$^3$. There are four examples in the Milky Way Galaxy: the
Arches and Quintuplet clusters near the Galactic center, and NGC\,3603
and Westerlund 1 in the disk.  Table\,\ref{Tab:observed} summarizes
the observed parameters for some known \YDCs.  In addition to the
clusters just mentioned, we also include R\,136, the central cluster
in the 30\,Doradus region, and MGG\,11, in the starburst galaxy M82,
which may contain an intermediate-mass black hole
\citep{2004Natur.428..724P}.

Motivated by these observations, as well as by theoretical
considerations, we distinguish two families of \YDC: those which are
effectively isolated, and those which are strongly perturbed by the
tidal field of their parent Galaxy.  We currently know only two
clusters in the latter category---the Arches and Quintuplet systems.
The other clusters in Table\,\ref{Tab:observed} are relatively
unperturbed.  In this paper we focus on the dynamics of the primordial
binary population in unperturbed young dense clusters.

The binary frequency and the distribution functions of binary orbital
parameters are poorly constrained by observations.  For two {\YDCs} we
have a rough estimate of the binary fraction: NGC\,3603, with an
estimated binary fraction of about 0.3 \citep{2004AJ....128..765S}, and
R\,136, which may contain up to 70\% binaries among its most massive
stars \citep{2002ApJ...574..762P}. These indications of binary
frequencies are based on the X-ray signatures of these clusters.
Further evidence for a high binary fraction in two star clusters near
the Galactic center is presented by \citep{2003ApJ...598..325Y}.

\begin{table*}[bp!]
\caption[]{\footnotesize Observed properties of selected young, dense
star clusters, mostly in or near the Milky Way.  Mass ($M$), age and
half-mass radius ($r_{\rm hm}$) are taken from the literature.  The
last two columns give the suspected binary fractions and an estimate
of the initial mass function.  Two question marks for the binary
fraction indicates that binaries are likely to be present but it is
impossible at this stage to give a reliable estimate, but within the
next few years better estimates may be expected. A horizontal dash
indicates that nothing is known about the binary fraction in that
cluster.  The mass function is designated as a Power-law (Pl) with the
exponent between parenthesis, or as a Salpeter function (identified by
S), a power-law with exponent 2.35, with lower and upper limits in
{\msun} as sub- and superscript, respectively.  }
\begin{flushleft}
\begin{tabular}{ll|lclcl} \hline
Name   &ref&   $M$& age & $r_{\rm hm}$ (pc)& $f_{\rm bin}$ & MF \\
       &  &[$10^4$\msun]& [Myr] & \\ \hline
Arches      &a& 60~~~~~ & 1--4~~~~~ & 0.23~ & ?? & Pl$^{100}_{8}(1.7)$\\
Quintuplet  &b& 15~~~~~ & 3--5~~~~~ & 0.5~~ & ?? & S$^{100}_{1}$ \\
\hline 
NGC\,3603   &c& 20~~~~~ & 2--3~~~~~ & 0.78~ & 0.3 & S$^{100}_{1}$ \\
Westerlund~1&d& 30~~~~  & 5--12~~~~ & 0.2~~ & ?? & S$^{100}_{8}$ \\
R\,136      &e& 50~~~~~ & 2--3~~~~~ & 0.5~~&$\sim 70$&S$^{100}_{8}$ \\
\hline
MGG-11      &f& 320~~~~~& 7--12~~~~ & 1.2$^\star$~& -- &S$^{15}_{1}$ \\
\hline
\end{tabular} \\
\medskip\footnotesize
References:
a) Figer et al.~(1999;2002);\nocite{2002ApJ...581..258F}\nocite{1999ApJ...514..202F}
b) Glass et al.~(1987);\nocite{1987MNRAS.227..373G}
c) Brandl (1999), \nocite{2004AJ....128..765S};
d) Vrba et al.~(2000); \nocite{1998A&AS..127..423P,2000ApJ...533L..17V}
e) Brandl et al.~(1996), \nocite{2002ApJ...574..762P};
f) McCrady et al. (2003)\\
\end{flushleft}
\label{Tab:observed}
\end{table*}

\section{The simulation model and initial conditions}

As initial conditions for our simulations we use a cluster of 131072
stars drawn from a Salpeter initial mass function between 1 and
100\,\msun. The lower mass limit is motivated by the observed age and
mass-to-light ratio of the young star cluster MGG11
\citep{2003ApJ...596..240M}.  A randomly selected subset of 13107
(10\%) stars receives a secondary companion with mass between
1\,\msun\, and the mass of the selected (primary) star.  Orbital
separations range from Roche-lobe contact (at about $E \simeq
10^3\,kT$, where $\frac32kT$ is the mean stellar kinetic energy) to a
binding energy of $E = 10$\,kT (at an orbital period of about a week),
while initial eccentricities are assumed to follow a thermal
distribution (see the left panel in
Figure\,\ref{Fig:Binaryparameters}).  The initial half-mass radius of
our simulation is 3.2\,pc, and even though our simulated cluster is
isolated we choose a King model with $W_0=12$ as the initial density
profile.

We solve the equations of motion of all the stars in the cluster, at
the same time calculating the evolution of the stars and binaries
using the Starlab\footnote{Starlab is available at {\tt
http://www.manybody.org/starlab/starlab.html}} software environment
\citep{2001MNRAS.321..199P}. We use the GRAPE-6 special-purpose
computer \citep{2003PASJ...55.1163M} to speed up the calculations.
Even though no older siblings of the {\YDCs} listed in
Table\,\ref{Tab:observed} are known, our cluster simulations generally
continue to ages of about 100\,Myr.

\begin{figure}[htbp!]
\begin{minipage}[b]{0.50\linewidth}
\psfig{figure=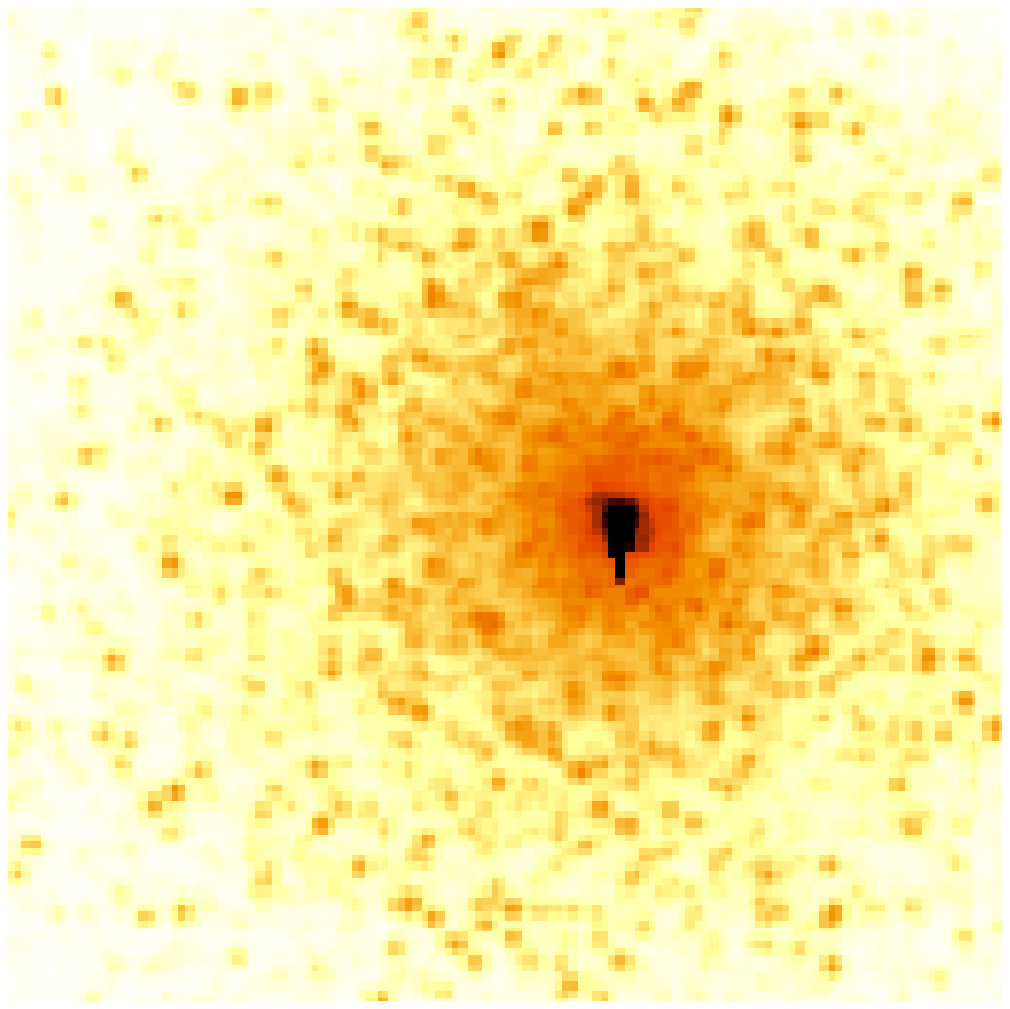,width=\linewidth,angle=0}
\end{minipage}\hfill
\begin{minipage}[b]{0.50\linewidth}
\psfig{figure=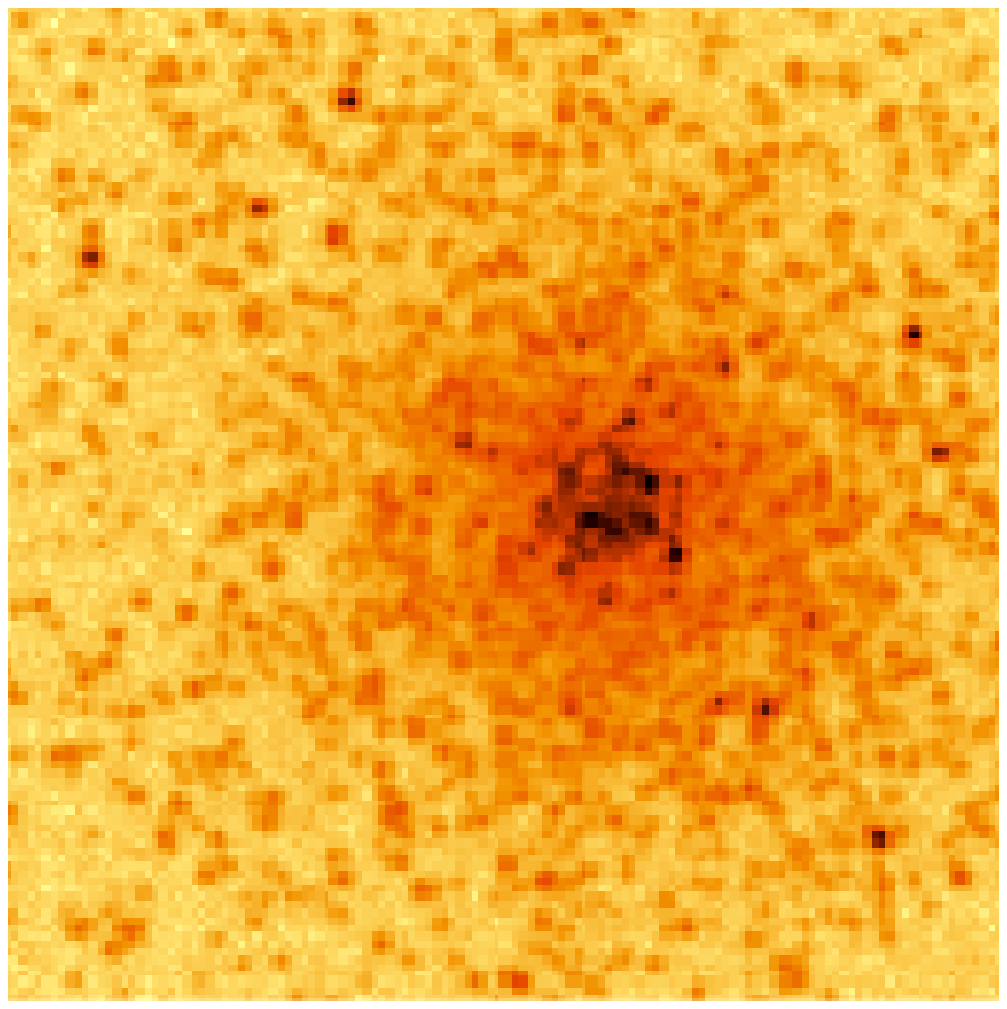,width=\linewidth,angle=0}
\end{minipage}\hfill
\caption[]{Simulated HST NICMOS images of our model cluster at a
distance of 8\,kpc, at birth (left) and at an age of about 115\,Myr
(right).  Images are taken with identical observation time in the
$I$-band; the image field of view is about 10\,pc across.}
\label{Fig:SOS}
\end{figure}

Rather than show the usual evolution of the core and Lagrangian radii
we show instead in Figure\,\ref{Fig:SOS} two simulated images of the
model cluster, one near the start of the simulation, and one near the
end.  Clearly, the cluster expands considerably with time.  As
discussed in \citep{2004astro.ph..6550Z}, the global evolution is
characterized by three quite distinct stages: {\bf A} an early
relaxation-dominated phase, followed by phase {\bf B}, during which
the $\sim 1$\,\%\, (by number) most massive stars evolve quickly and
lose an appreciable fraction of their mass.  Finally, in phase {\bf
C}, stellar evolution slows and relaxation takes over until the
cluster dissolves \citep{2004astro.ph..6550Z}.  For this particular
simulation, phase {\bf A} lasts for about 3\,Myr, phase {\bf B} lasts
until $\sim50$\,Myr, followed by phase {\bf C} for the remainder of
the run.

\section{Results}

\begin{figure}[htbp!]
\psfig{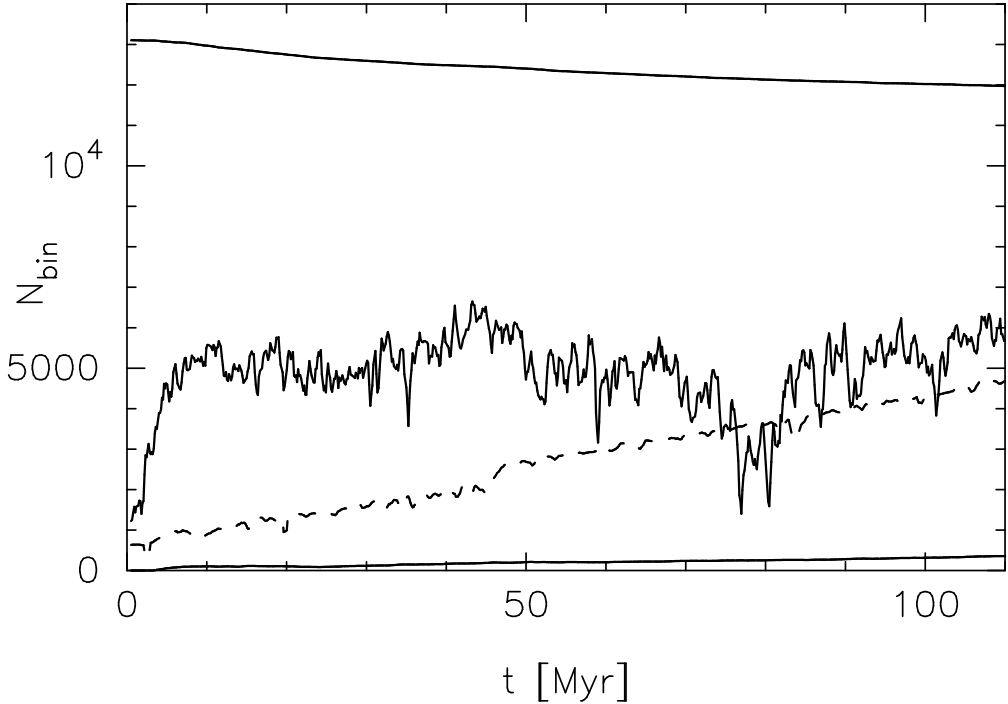}
\caption[]{Evolution of the number of hard binaries including binaries
softer than 10\,kT for our simulation (top solid curve).  Binaries
with binding energy $>100$\,kT are presented as the dashed curve, and
the number of hard binaries within 10\,\rcore\, are indicated by the
middle solid curve. The lower solid curve shows the binaries
containing at least one compact object---black hole, neutron star or
white dwarf.  An enlargement of this curve is presented in
Fig.\,\ref{Fig:rnmBinaryFraction}. }
\label{Fig:TotalBinaryFraction}
\end{figure}

Figure\,\ref{Fig:TotalBinaryFraction} shows the evolution of the
number of binaries in our simulated cluster. The total number of
binaries drops by $\sim 9$\% in 115\,Myr, whereas the number of stars
drops by only 1\%. The total cluster mass drops by about 40\% during
the same period.

The top solid curve gives the total number of binaries throughout the
simulations, from birth to an age of 115\,Myr. The lower solid curve
gives the number of binaries within 10\,\rcore. We use ten times the
core radius (\rcore) here because the number of binaries withing one
core radius is so small. The dashed curve shows the number of very
hard binaries, having binding energy $E > 100$\,kT.

\begin{figure}[htp!]
\psfig{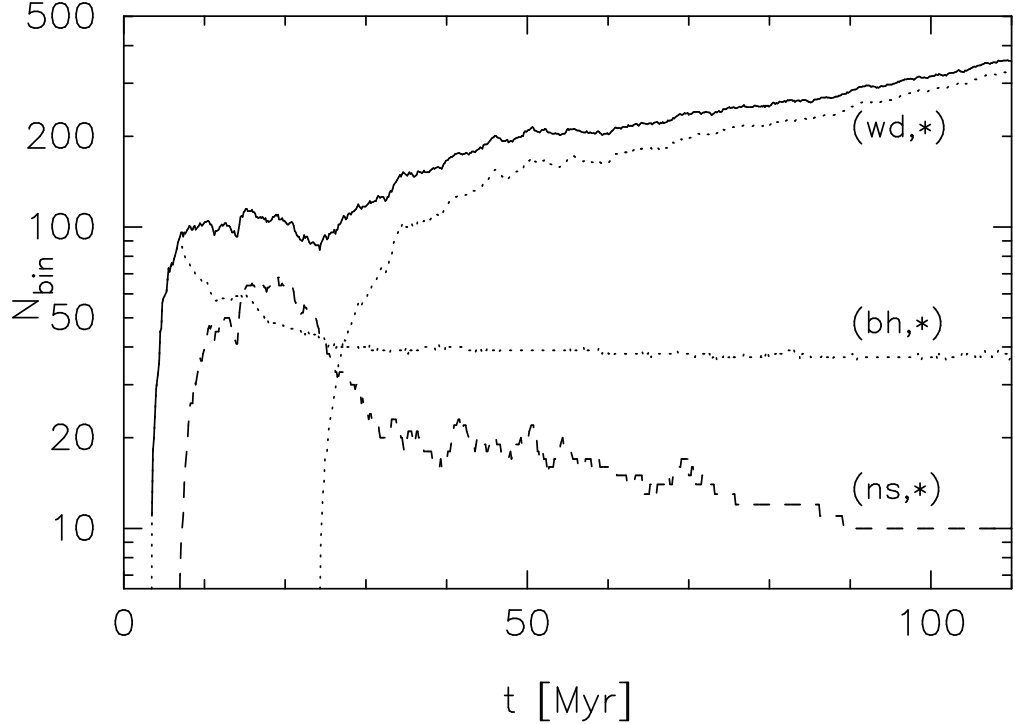}
\caption[]{Magnification of the evolution of the number of compact
objects in the simulation.  The solid curve shows the number of
binaries containing at least one black hole, neutron star or white
dwarf (identical to the bottom solid curve in
Figure\,\ref{Fig:TotalBinaryFraction}). The dashed curve gives the
number of binaries with at least one neutron star, indicated as (ns,
$\star$).  The two dotted curves, indicated as (bh, $\star$) and (wd,
$\star$), give the number of binaries containing at least one black
hole and at least one white dwarf, respectively.}
\label{Fig:rnmBinaryFraction}
\end{figure}

In Figure\,\ref{Fig:rnmBinaryFraction} we present, as functions of
time, the number of binaries with at least one compact object, black
hole, neutron star or white dwarf.  The number of black holes in
binaries rises sharply shortly after the start of the simulation with
a peak near 8\,Myr. This is the moment when the turn off drops below
the minimum mass ($\sim 20$\,\msun) for forming black holes:
lower-mass stars form neutron stars.  This threshold is also visible
as the sharp increase of the number of binaries containing a neutron
star. The number of black holes in binaries drops rapidly from this
moment on, because many of their companions form neutron stars in
supernova explosions.  These newly formed neutron stars receive a much
higher asymmetric kick velocity during their formation
\citep{1994Natur.369..127L} than black holes
\citep{2004astro.ph..7502G}. Note that binaries containing a neutron
star and a black hole are counted twice in this figure, once among the
(bh, $\star$) and once for (ns, $\star$).

The middle solid curve in Figure\,\ref{Fig:rnmBinaryFraction} shows an
interesting dip between about 75\,Myr and 80\,Myr binaries within
{10\,\rcore} are depleted.  The explanation for this local depression
is the occurrence of core collapse.  This is rather surprising, as the
energy stored in primordial binaries is sufficient to support the
cluster core and to prevent core collapse.

For a star cluster with a broad mass spectrum and without primordial
binaries, core collapse occurs in about 20\% of the initial half-mass
relaxation time \citep{2003ApJ...596..314M}. With a half-mass
relaxation time of $\sim 360$\,Myr in our simulation, one therefore
expects core collapse in about 70\,Myr.  We attribute the fact that
the observed collapse happens about 10\,Myr later than that to the
effect of stellar mass loss, as by this time the cluster has lost
about 36\% of its mass by stellar evolution.  (Note that less mass
would have been lost by the cluster if we had adopted a lower limit to
the initial-mass function.)  Most of the mass loss happens in the
cluster core, effectively driving an adiabatic expansion; it appears
that binary heating does not play an important role in the core
evolution.

Even though we have not presented here definitive proof that the
binary population is ineffective in influencing this phase of the core
evolution, the collapse of the cluster core suggests that binary
heating delays core collapse by at most 10\,Myr.  This may be
contrasted with the results of earlier Monte-Carlo simulations, which
indicate that core collapse is significantly delayed by the presence
of a rich population of primordial binaries
\citep{2003ApJ...593..772F}.

The first white dwarfs appear at 23.8\,Myr,
at a turn-off mass of about 10\,\msun. Isolated stars
(i.e.\,unperturbed by any companion) are expected to evolve into white
dwarfs if their initial mass is less than roughly 8\,\msun.  However,
in binary systems, early stripping of the hydrogen envelope may cause
more massive stars also to become white dwarfs instead of collapsing
to neutron stars.  The population of compact binaries of clusters
older than about 40\,Myr is dominated by white dwarfs, although at a
formation rate of about 100 white-dwarf binaries per 30\,Myr, it will
take about 2\,Gyr before they become more common than primordial
main-sequence binaries.

\begin{figure}[htbp!]
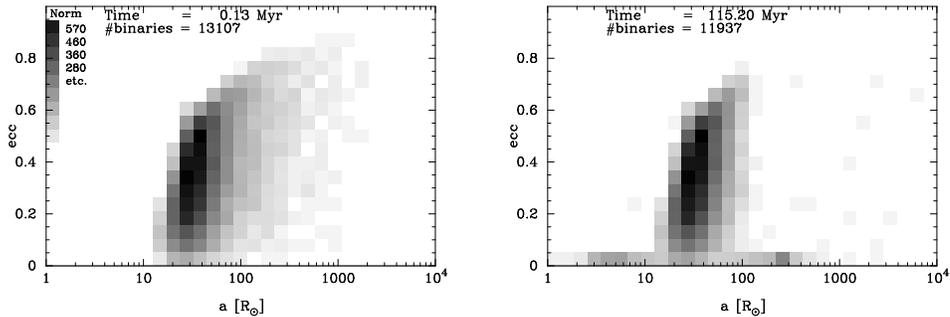

\begin{minipage}[b]{0.50\linewidth}
\psfig{figure=./MGG11.N128k.W12.r3.2.SFM.B0.1.T0Myr.AEall.ps,width=\linewidth,angle=-90}
\end{minipage}\hfill
\begin{minipage}[b]{0.50\linewidth}
\psfig{figure=./MGG11.N128k.W12.r3.2.SFM.B0.1.T100Myr.AEall.ps,width=\linewidth,angle=-90}
\end{minipage}\hfill
\caption[]{Distribution of the orbital separation (in \rsun) and
eccentricity of all binaries in our simulation near zero age (left
frame) and at the end of the calculation, at about 115\,Myr (right
frame). Darker shades indicate more binaries, as is indicated in the
normalization in the upper-left corner of the left panel.  The total
number of binaries and the age of the cluster are indicated near the
top of each frame.  }
\label{Fig:Binaryparameters}
\end{figure}

Figure\,\ref{Fig:Binaryparameters} shows the distribution in
semi-major axis and eccentricity of the binary population at the start
of our simulation (left panel) and at 115\,Myr (right). We have
rendered the image in gray-shades, with darker shades indicating more
systems within the bin. The left panel in Figure
\ref{Fig:Binaryparameters} represents essentially the zero-age
population, except that we show the system after about one cluster
crossing time (0.1\,Myr) to eliminate (for example) any spurious
initializations of eccentric binaries which should circularize shortly
after formation. The orbital separation ranges from about 15\,\rsun\,
to about 1000\,\rsun.  The lack of very high eccentricities ($e\apgt
0.9$) is mainly due to tidal effects within binary systems.  An
encounter is unlikely to ionize a binary, but effectively changes its
eccentricity; highly eccentric binaries ($e \apgt 0.6$) typically
become more circular after an encounter.

In the right panel of Figure\,\ref{Fig:Binaryparameters} we show the
distribution in semi-major axis and eccentricity for the binaries
remaining in our simulation at an age of 115\,Myr.  The two most
striking differences compared to the left (zero age) panel are the
depletion of binaries with orbital separations $a \apgt 200$\,\rsun,
and the enormous increase in binaries on circular orbits.

The pile-up in Figure\,\ref{Fig:Binaryparameters} of wide ($a\apgt
200$\,\rsun) binaries with circular orbits and the depletion of wide
eccentric binaries is caused mainly by the internal evolution of the
binaries in combination with the initial conditions.  The initial
binary separations were selected between Roche-lobe contact and a
binding energy of $E = 10$\,kT.  The hard-soft boundary for a binary
with orbital separation $a$ and total mass $m$ may be estimated as
\begin{equation}
  a \approx N \left( m/M \right)^2 R.
\end{equation}
Here $N$, $M$, and $R$ are the number of stars and the total cluster
mass and half-mass radius.  Binaries with total mass exceeding
5\,{\msun} will, according to our initial requirement that all
binaries have binding energy $E > 10$\,kT, have orbital separations
smaller than 200\,\rsun.  The turnoff mass for our cluster at an age
of 115\,Myr is about 4.5\,\msun, so binary components more massive
than that are either ascending the giant branch or have already done
so. The binaries with initial orbital separations exceeding $\sim
200$\,\rsun\, host mainly rather massive stars, which have already
evolved off the main sequence by the time the cluster is 115\,Myr old,
and therefore these binaries have experienced strong internal
evolution. Note here that wide binaries tend to shrink upon Roche-lobe
overflow, due to angular momentum loss via non-conservative mass
transfer, whereas shorter-period binaries tend to expand; this leads
to a pile-up of circular binaries around $a = 100$--300\,\rsun.
Binaries containing lower-mass stars generally have tighter orbits,
and do not populate this region, at least not during the cause of the
simulation.

The few eccentric binaries with orbital separations $a \apgt
200$\,{\rsun} at an age of 115\,Myr have experienced two supernovae;
in most cases (11 out of 14) both supernovae led to the formation of a
black hole.  The few binaries to the left of the circularization
boundary are in the process of rapid circularization due to tidal
effects.

\section{Summary}

We report the results of a large (128k) direct N-body simulation with
10\% hard ($E\apgt10$\,kT) primordial binaries.  Our simulations
include the effects of stellar mass loss, collisions, and coalescence,
and the internal evolution of binaries.

Our main conclusions may be summarized as follows: core collapse in
the simulation occurs at an age of about 80\,Myr. In a star cluster
with similar initial conditions but without stellar mass loss and
without primordial binaries, core collapse would be expected to occur
at about 70\,Myr.  We therefore argue that core collapse is slightly
delayed by stellar mass loss and the presence of the binaries.  The
former appears to be the main cause of the delayed core collapse.

The number of binaries containing one or two compact objects increases
steadily from the start of the simulation.  After about 30\,Myr, the
number of binaries with at least one black hole remains roughly
constant at about 40, whereas the number of binaries containing at
least one white dwarf continues to grow at a roughly constant rate of
about 100 per 30\,Myr.


\section*{Acknowledgment}
We are grateful to Jun Makino and Piet Hut for numerous discussions,
and in particular to Makino for the use of his GRAPE-6, on which many
of our simulations have been performed. Additional simulations are
carried out using the GRAPE-6 systems at Drexel University and at the
MoDeStA computer at the University of Amsterdam.  This work was
supported by NASA ATP grant NAG5-10775, the Royal Netherlands Academy
of Sciences (KNAW), the Dutch organization of Science (NWO), and by
the Netherlands Research School for Astronomy (NOVA).


\begin{thebibliography}{}

\bibitem[\protect\astroncite{{Figer} et~al.}{1999}]{1999ApJ...514..202F}
{Figer}, D.~F., {McLean}, I.~S., {Morris}, M. 1999, \apj, 514, 202

\bibitem[\protect\astroncite{{Figer} et~al.}{2002}]{2002ApJ...581..258F}
{Figer}, D.~F., {Najarro}, F., {Gilmore}, D., {Morris}, M., {Kim}, S.~S.,
  {Serabyn}, E., {McLean}, I.~S., {Gilbert}, A.~M., {Graham}, J.~R., {Larkin},
  J.~E., {Levenson}, N.~A., {Teplitz}, H.~I. 2002, \apj, 581, 258

\bibitem[\protect\astroncite{{Fregeau} et~al.}{2003}]{2003ApJ...593..772F}
{Fregeau}, J.~M., {G{\" u}rkan}, M.~A., {Joshi}, K.~J., {Rasio}, F.~A. 2003,
  \apj, 593, 772

\bibitem[\protect\astroncite{{Glass} et~al.}{1987}]{1987MNRAS.227..373G}
{Glass}, I.~S., {Catchpole}, R.~M., {Whitelock}, P.~A. 1987, \mnras, 227, 373

\bibitem[\protect\astroncite{{Gualandris} et~al.}{2004}]{2004astro.ph..7502G}
{Gualandris}, A., {Colpi}, M., {Portegies Zwart}, S.~F., {Possenti}, A. 2004, ApJ in press, (astro-ph/0407502)

\bibitem[\protect\astroncite{{Lyne} \& {Lorimer}}{1994}]{1994Natur.369..127L}
{Lyne}, A.~G., {Lorimer}, D.~R. 1994, \nat, 369, 127

\bibitem[\protect\astroncite{{Makino} et~al.}{2003}]{2003PASJ...55.1163M}
{Makino}, J., {Fukushige}, T., {Koga}, M., {Namura}, K. 2003, \pasj, 55, 1163

\bibitem[\protect\astroncite{{McCrady} et~al.}{2003}]{2003ApJ...596..240M}
{McCrady}, N., {Gilbert}, A.~M., {Graham}, J.~R. 2003, \apj, 596, 240

\bibitem[\protect\astroncite{{McMillan} \& {Portegies
  Zwart}}{2003}]{2003ApJ...596..314M}
{McMillan}, S.~L.~W., {Portegies Zwart}, S.~F. 2003, \apj, 596, 314

\bibitem[\protect\astroncite{{Piatti} et~al.}{1998}]{1998A&AS..127..423P}
{Piatti}, A.~E., {Bica}, E., {Claria}, J.~J. 1998, \aaps, 127, 423

\bibitem[\protect\astroncite{{Portegies Zwart}}{2004}]{2004astro.ph..6550Z}
{Portegies Zwart}, S.~F. 2004, 
Book of the Como School of Physics, {\em Joint Evolution of Black Holes and Galaxies}, Graduate School in Contemporary Relativity and Gravitational Physics, May 05-10, 2003 in Como, Italy, 
(astro-ph/0406550)

\bibitem[\protect\astroncite{{Portegies Zwart}
  et~al.}{2004}]{2004Natur.428..724P}
{Portegies Zwart}, S.~F., {Baumgardt}, H., {Hut}, P., {Makino}, J., {McMillan},
  S.~L.~W. 2004, \nat, 428, 724

\bibitem[\protect\astroncite{{Portegies Zwart}
  et~al.}{2001}]{2001MNRAS.321..199P}
{Portegies Zwart}, S.~F., {McMillan}, S.~L.~W., {Hut}, P., {Makino}, J. 2001,
  \mnras, 321, 199

\bibitem[\protect\astroncite{{Portegies Zwart}
  et~al.}{2002}]{2002ApJ...574..762P}
{Portegies Zwart}, S.~F., {Pooley}, D., {Lewin}, W.~H.~G. 2002, \apj, 574, 762

\bibitem[\protect\astroncite{{Stolte} et~al.}{2004}]{2004AJ....128..765S}
{Stolte}, A., {Brandner}, W., {Brandl}, B., {Zinnecker}, H., {Grebel}, E.~K.
  2004, \aj, 128, 765

\bibitem[\protect\astroncite{{Vrba} et~al.}{2000}]{2000ApJ...533L..17V}
{Vrba}, F.~J., {Henden}, A.~A., {Luginbuhl}, C.~B., {Guetter}, H.~H.,
  {Hartmann}, D.~H., {Klose}, S. 2000, \apjl, 533, L17

\bibitem[\protect\astroncite{{Yusef-Zadeh}}{2003}]{2003ApJ...598..325Y}
{Yusef-Zadeh}, F. 2003, \apj, 598, 325

\end{thebibliography}
\end{document}